\documentclass[12pt]{article}
\usepackage{amsmath,amssymb,epsfig,amsfonts,graphicx,euscript}
\usepackage[pdftex,bookmarks,bookmarksnumbered,linktocpage,pdfstartview=FitH]{hyperref}
\hypersetup{colorlinks,%
citecolor=red,%
filecolor=blue,%
linkcolor=blue,%
urlcolor=blue,%
pdftex}
\usepackage[all]{hypcap}
\numberwithin{equation}{section}
\usepackage{cite}
\newcommand{\bse}{\begin{subequations}}
\newcommand{\ese}{\end{subequations}}
\newcommand{\be}{\begin{equation}}
\newcommand{\ee}{\end{equation}}
\newcommand{\bea}{\begin{eqnarray}}
\newcommand{\eea}{\end{eqnarray}}
\newcommand{\ba}{\begin{array}}
\newcommand{\ea}{\end{array}}
\newcommand{\h}{\frac{1}{2}}
\newcommand{\G}{{\tilde{g}}}

\begin{document}
\hfill%
\vbox{
    \halign{#\hfil        \cr
           IPM/P-2012/010\cr
                     }
      }
\vspace{1cm}
\begin{center}
{ \Large{\textbf{Meson Thermalization in Various Dimensions}}} 
\vspace*{2cm}
\begin{center}
{\bf Mohammad Ali-Akbari\footnote{aliakbari@theory.ipm.ac.ir}, Hajar Ebrahim\footnote{hebrahim@ipm.ir}}\\%
\vspace*{0.4cm}
{\it {${}^1$School of Particles and Accelerators,\\ ${}^2$School of Physics,\\ Institute for Research in Fundamental Sciences (IPM),\\
P.O.Box 19395-5531, Tehran, Iran}}  \\

\vspace*{1.5cm}
\end{center}
\end{center}

\vspace{.5cm}
\bigskip
\begin{center}
\textbf{Abstract}
\end{center}
In gauge/gravity duality framework the thermalization of mesons in strongly coupled $(p+1)$-dimensional gauge theories is studied for a general D$p$-D$q$ system, $q\geq p$, using the flavour D$q$-brane as a probe. Thermalization corresponds to the horizon formation on the flavour D$q$-brane. We calculate the thermalization time-scale due to a time-dependent change in the baryon number chemical potential, baryon injection in the field theory. We observe that for such a general system it has a universal behaviour depending only on the t'Hooft coupling constant and the two parameters which describe how we inject baryons into the system. We show that this universal behaviour is independent of the details of the theory whether it is conformal and/or supersymmetric.

\newpage

\tableofcontents

\section{Introduction and Results}

One of the interesting experiments in particle physics is the heavy ion collision done at RHIC (Relativistic Heavy Ion Collider) and LHC (Large Hadron Collider). In these experimental set-ups two pancakes of heavy nuclei such as Gold(Au) or Lead(Pb) are collided at a relativistic speed. Very soon after the collision a new phase of matter called QGP (Quark-Gluon Plasma) is produced. Understanding the properties of the QGP has attracted a lot of attention. It's been realized through hydrodynamic simulations and experimental observations that the plasma is strongly coupled with very low viscosity over entropy density, $(\frac{\eta}{s})$ \cite{Shuryak:2003xe,Shuryak:2004cy}.

A very challenging observation is the very rapid thermalization of plasma. At almost 1fm/c after the collision the plasma reaches local equilibrium (thermalizes) and its dynamics is approximately described by ideal fluid hydrodynamics \cite{Heinz:2004pj,Luzum:2008cw}. Since the system is strongly coupled and perturbative calculations can not be applied, it is difficult to study an out-of-equilibrium process such as thermalization. Models of QCD such as lattice QCD have not been able to achieve much to explain this phenomenon.

Recently AdS/CFT correspondence \cite{Maldacena}, or more broadly, the gauge/gravity duality has gained some success in making toy models  to describe this process.
Generally speaking, the AdS/CFT duality  has provided a
powerful framework to study strongly coupled gauge theories.
According to the original statement of the correspondence, type IIB string theory on
$AdS_5\times S^5$ is dual to $D=4$, ${\cal{N}}=4$, $SU(N)$ super
Yang-Milles (SYM) theory. This idea was then generalized to a broader class of
gauge/gravity dualities in various dimensions \cite{Itzhaki}. It was
shown that in the large $N$ limit, the strongly coupled $SU(N)$ gauge theory living on the worldvolume of $N$ coincident D$p$-branes is dual to the supergravity on the
near horizon geometry of the D$p$-branes. Although for $p=3$ the gauge theory enjoys conformal symmetry, the gauge coupling is dimensionful for the other
values of $p$.

The absence of conformal invariance in the gauge theory leads to the
radial variation of both the string coupling (dilaton) and the spacetime
curvature in the dual description. This is due to the fact that duality relates the radial
coordinate transverse to the D$p$-brane to the energy scale in the
gauge theory. The supergravity background is only
reliable for weak string coupling and small curvature. This
is provided in a regime of energy where \cite{Itzhaki} %
\be %
 1\ll g_{eff}\ll N^{\frac{4}{7-p}}~.
\ee %
$g_{eff}$ is the dimensionless effective coupling,
$g_{eff}=g_{YM}^2Nu^{p-3}$, where $u$ is the radial coordinate. In this intermediate regime of energy
the dual gauge theory is always strongly coupled.

AdS/CFT techniques have been applied to study rapid thermalization observed at RHIC \cite{Chesler:2008hg,Bhattacharyya:2009uu,Ebrahim:2010ra,Das:2010yw,Hashimoto}.
The thermalization in field theory which is an out-of-equilibrium process happens after an injection of energy into the system. In the gauge/gravity context this injection can be done in two different ways: by directly adding a time-dependent source to the boundary field theory which corresponds to turning on a non-normalizable mode \cite{Chesler:2008hg,Bhattacharyya:2009uu} or by introducing a time-dependent coupling in the field theory which is represented by nontrivial time-dependent classical solutions of some kinds of probe branes \cite{Das:2010yw,Hashimoto}. In the first method the source is assumed to be nonzero only for a limited time interval. In the bulk gravity dual this corresponds to the collapse of matter and formation of the horizon; in other words {\it{black hole formation}}. In \cite{Chesler:2008hg} the authors have been able to quantitatively give an approximate value for the isotropization (thermalization) time of the plasma which is close to the experimental result. In the other method the horizon formation happens on the probe brane and then the energy can flow from the probe brane to the bulk gravitational degrees of freedom. This corresponds to the dissipation of energy into the field theory.

Note that in the first method the field theory is pure YM and conformal. While in the other one the field theory contains fundamental matters and mesons in addition to the pure YM. Studying the thermalization in the meson sector is the subject of an interesting work \cite{Hashimoto} where the authors have modeled the thermalization by baryon injection into the system. This idea comes from the observed sudden change in the baryon chemical potential at QGP production.

The addition of fundamental matter to the SYM theory which lives on the boundary of AdS space is done by adding the flavour branes to the dual gravity system. This has been proposed for the first time in \cite{Karch:2002sh} where $N_f$ D7-branes are added to the ${\rm{AdS}}_5\times {\rm{S}}^5$ background in the probe limit where $\frac{N_f}{N}$ is small. This will reduce the number of supersymmetries of the original theory from 16 to 8 and will add a matter hypermultiplet in the fundamental representation to the SYM theory. The hypermultiplet describes the dynamical quarks living in four dimensions and the fluctuations of the flavour brane explain the meson spectrum of the field theory. In another interesting work this system has been extended to more general settings of D$p$-D$q$ brane configurations \cite{Myers}.

In the D3-D7 brane set-up the authors of \cite{Hashimoto} have studied thermalization in the meson sector of the SYM theory. The thermalization is modeled by a sudden change in the baryon number. In the gravity dual description this change is realized by throwing the baryonically-charged fundamental strings from the boundary to the bulk. The end points of the strings stretched between D3 and D7-branes act as the source for the gauge field on the D7-brane. Therefore this set-up provides a time-dependent gauge field configuration. Using DBI action this induces a time-dependent metric and eventually emergence of an apparent horizon on the D7-brane. Hence the baryon injection leads to the horizon formation on the D7-brane which is the signal of thermalization in the meson sector of the field theory.

An intriguing observation of \cite{Hashimoto} is that the
thermalization time in the field theory for the D3-D7 system can be
written only in terms of a few parameters which are: the gauge theory
t'Hooft coupling $(\lambda=g_{YM}^2N)$, maximum baryon number
density$(n_B)$ and the inverse of the variation time-scale of the
baryon number $(\omega)$. Note that $n_B$ and $\omega$ describe how the baryon charge density changes in the system. Therefore the thermalization time  is independent of
the details of the theory by which we mean the form of the Lagrangian. The authors generally argue that this possible
universal behaviour might be generalized to other gauge theories. In
this paper we would like to examine this idea and generalize their
computation to the D$p$-D$q$ system.

Interestingly for such a general system where the gauge theory is not conformal, the background
metric is not AdS and as mentioned above gauge/gravity duality can be trusted only in an intermediate range of energy \cite{Itzhaki}, we observe
similar universal scaling behaviour. To be more explicit, by universality we mean the thermalization time of the form
\be%
\label{universal}
t_{th}\sim \bigg{(}\frac{\lambda^\alpha}{n_B^2 ~\omega^{2}}\bigg{)}^{\beta}~,
\ee%
where as we will see in the following sections $\alpha$ and $\beta$
are fixed in terms of $p$ using the equation of the apparent
horizon. 

The thermalization time is obtained by studying the dynamics of the
scalar mesons in D$p$-D$q$ system after the time-dependent baryon
injection. This can be done by investigating the fluctuations of the
transverse and parallel directions of the D$q$-brane to the
D$p$-brane. We will show that although the equations of the apparent
horizon are different in these two cases, the thermalization
time-scales are yet the same.

We will also observe that even when the background is not
supersymmetric, which means $q\neq p,~p+2,~p+4$, this behaviour
still persists. The only change in \eqref{universal} occurs in
$\alpha$ and $\beta$. These results confirm, to some extent, the
claim of \cite{Hashimoto} that the thermalization time-scale shows a
universal behaviour.

\section{Review on D$p$-D$q$ System}
In the introduction we mentioned that we are interested in studying
the thermalization of mesons in strongly coupled field theories in
the context of gauge/gravity duality. Therefore in this section we will give a
brief review on D$p$-D$q$ brane configurations that are used to
explain the meson spectrum in the gauge theory.

The supergravity
solution corresponding to near horizon limit of $N$
coincident D$p$-branes is \cite{Itzhaki} %
\be\label{background}\begin{split} %
 ds^2&=H^{-\h}(-dt^2+dx_p^2)+H^\h(du^2+u^2d\Omega_{8-p}^2), \cr%
%
 d\Omega_{8-p}^2&=d\theta^2+\sin^2\theta d\Omega_k^2+\cos^2\theta
 d\Omega^2_{7-p-k},
\end{split}\ee %
where %
\be\begin{split} %
 H(u)&=(\frac{R}{u})^{7-p},\ \ e^\phi=H^{\frac{3-p}{4}},\ \
 C_{01...p}=H^{-1}, 
\end{split}\ee %
written in the string frame. The Dilaton field is represented by $\phi$
and $C_{01...p}$ is a $(p+1)$-form field coupled to D$p$-branes. The
length scale $R$ is defined in terms of the string length scale
$l_s=\sqrt{\alpha'}$ and the string coupling $g_s = e^{\phi_{\infty}}$
\be %
 R^{7-p}=(4\pi)^{\frac{5-p}{2}}\ \Gamma({\frac{7-p}{2}})\ g_sN\
 l_s^{7-p}.
\ee %
It is easy to see that for the special case $p=3$,
\eqref{background} reduces to $AdS_5\times S^5$. We are considering
$p\leq 4$ cases because there is no decoupling limit for $p\geq 5$
and hence no dual gauge theory \cite{Itzhaki}.

According to gauge/gravity duality, a strongly coupled
$SU(N)$ SYM theory living on the $(p+1)$-dimensional worldvolume of $N$ coincident
D$p$-branes is dual to the supergravity on the above background \eqref{background} in the large $N$ limit
\cite{Itzhaki}. The isometry group of this background geometry is
$SO(1,p)\times SO(9-p)$ which in the dual gauge theory corresponds to
space-time Lorentz symmetry and R-symmetry, respectively. The gauge theory coupling constant is related to the string coupling and the
string length through
\be %
 g_{YM}^2=2\pi g_s(2\pi l_s)^{p-3}. %
\ee %

One can add a stack of $N_f$ D$q$-branes to the above background and
study it in the probe limit where $\frac{N_f}{N}$ is small. The low
energy effective action for a D$q$-brane in an arbitrary
background is \cite{Myers2}%
\be\begin{split}\label{action} %
 S &= S_{\rm{DBI}}+ S_{\rm{CS}}~,\cr
 S_{{\rm{DBI}}}&=-\tau_q\int d^{q+1}\xi\
 e^{-\phi}\sqrt{-\det(g_{ab}+B_{ab}+2\pi\alpha'F_{ab})}~,\cr
 S_{\rm{CS}} &=\tau_q\int P[\Sigma C^{(n)}e^B]e^{2\pi\alpha'F}~,
\end{split}\ee %
where induced metric $g_{ab}$ and Kalb-Ramond field $B_{ab}$ are given by %
\be\begin{split} %
 g_{ab}=G_{MN}\partial_a X^M\partial_b X^N,\cr
 B_{ab}=B_{MN}\partial_a X^M\partial_b X^N.
\end{split}\ee %
$a,b,...$ indices are used to describe worldvolume coordinate running over
0,1,...,$q$. The capital indices $M,N,...$ are used to denote
space-time coordinates. $G_{MN}$ is the background metric that is
introduced in \eqref{background} and $F_{ab}$ is the field strength
of the gauge field living on the D$q$-brane. Note that in the background we consider here the B-field is zero. In the Chern-Simons part,
$C^{(n)}$ denotes the $(n+1)$-form Ramond-Ramond potential and $P[...]$ is
the pull-back of the bulk space-time tensors to the D$q$-brane
worldvolume. The D$q$-brane tension is %
\be %
 \tau_q=\frac{1}{(2\pi)^q\ l_s^{q+1}g_s}~.
\ee %

In order to embed a probe D$q$-brane in the background \eqref{background} one can introduce the following change of coordinates %
\be\begin{split} %
 \rho&=u\sin\theta~, \cr %
 \sigma&=u\cos\theta~.
\end{split}\ee %
Hence the background metric becomes %
\be %
\label{induced metric}
 ds^2=H^{-\h}(-dt^2+dx_p^2)+H^\h\Big(d\rho^2+\rho^2d\Omega_k^2+d\sigma^2+\sigma^2d\Omega^2_{7-p-k}\Big). %
\ee %
This configuration of the D$p$-D$q$ branes can be schematically shown as
\be %
           \begin{array}{cccccccccccc}\label{graph}
                & t & x_1 & .. & x_d & x_{d+1} & .. & x_p & \rho & \Omega_k & \sigma & \Omega_{7-p-k}  \\
             Dp & \times & \times & \times & \times & \times & \times &  \times &  &   &   &  \\
             Dq & \times & \times & \times & \times &  &  &  & \times & \times &  &

           \end{array}
\ee %
where %
\be %
 q=k+d+1.
\ee %
In order to get a stable background one has to consider the brane configurations which preserve supersymmetry. This condition dictates
that $q$ should be $p$, $p+2$, $p+4$ and correspondingly $k$ is $1$, $2$ and $3$. Note that this embedding breaks the initial
isometry group of the D$p$-brane to
\be %
 SO(1,d) \times SO(p-d) \times SO(k+1) \times SO(8-p-k).
 \ee%

Supersymmetrically embedding $N_f$ flavour D$q$-branes into the background obtained from the near horizon limit of $N$ D$p$-branes corresponds
to coupling $N_f$ flavour dynamical matter fields in the fundamental representation (dynamical quarks) to the SYM theory which lives on
the D$p$-brane. This reduces the number of supersymmetries of the original theory from 16 to 8. The fundamental matter fields which form a ${\cal {N}}=2$ hypermultiplet arise as the lightest modes of the strings stretched between D$p$ and D$q$-branes.

The spectrum of the fluctuations of a D$q$-brane ($N_f=1$) corresponds to the mesonic spectrum of a $SU(N)$ SYM theory in $p+1$ dimensions which is coupled to a hypermultiplet in the fundamental representation \cite{Myers}. Note that in the case where $q=p+4$, the fundamental fields live on the full $p+1$ dimensions of the gauge theory. But for $q=p+2$ and $q=p$ they are confined to a $(d+1)$-dimensional defect where the D$p$ and D$q$-branes coincide.

\section{Thermalization}
The thermalization in field theory is the result of the injection of energy into the system. One way to realize it in the context of gauge/gravity duality is to inject baryon charge to the strongly coupled field theory. This has been done for the first time in \cite{Hashimoto} where the dynamics of mesons or in other words the thermalization of mesons in D3-D7 system has been studied. In this section we will elaborate more on this process by generalizing it to D$p$-D$q$ system.

\subsection{Baryon injection}
The injection of baryons into the strongly coupled field theory is presented in the gauge/gravity context by throwing byronically
charged F-strings on the flavour D$q$-brane. The end points of the strings stretched between D$q$ and D$p$-branes resemble
quarks in the gauge theory. The quarks are assumed to be massless. They act as a source for the background gauge field on the flavour brane. The injection of
quarks into the system results in a sudden change in the baryon number chemical potential.

In the AdS/CFT correspondence,
static baryon chemical potential is identified with the time component of the gauge field in the following way
\bea%
\mu &=& \int_{0}^{\infty} du ~\partial_{u} A_t(u)~, \cr &=&
A_t(\infty) - A_t(0)~, \eea where $u$ represents the radial
direction in the ${\rm{AdS}}$ space\footnote{Note that this identification is done in the gauge where $A_u$ is zero.}. Therefore in order to explain
a time-dependent chemical potential which results from the
injection of quarks to the system we must add the source term
\be %
 S_{{\rm{current}}}=\int d^{q+1}\xi \sqrt{-g}A_aJ^a,
\ee %
to the action \eqref{action}. Note that the currents, $J^a$s, are now time-dependent and will produce time-dependent gauge fields.

Since the end points of the open strings move on the light geodesics,
without loss of generality, the
current components can be fixed to be $J^{\rho}$ and $J^t$. We assume that the massless quarks are moving along the null line $x^-$ defined as
\bea %
x^\pm &=& t \pm z = t \mp \int H^{\frac{1}{2}} d\rho~. 
\eea %
Hence the currents are only functions of $x^-$.

With the above assumptions the D$q$-brane DBI action for the background \eqref{induced metric} in the presence of the
source terms reduces to\footnote{Note that $\tau_q j^a=\frac{R^8}{z^5}J^a$. }
\be\begin{split} %
 S=&-\tau_q V_d{\rm{Vol}}(S^k)\int dt d\rho
 \Big[H^{\frac{p-q+2(k-1)}{4}}\rho^k\sqrt{1-(2\pi\alpha')^2F_{t\rho}^2}-A_tj^t-A_\rho
 j^\rho\Big].
\end{split}\ee %
To solve the equation of motion for the gauge field we can
consistently choose
\be\label{current}%
j^\rho = - H^{\frac{-1}{2}} j^t  =
g'(x^-),
\ee%
which satisfies the current conservation equation
\be %
 \nabla_a J^a=0.
\ee %
Therefore the classical time-dependent solution for the gauge field is
\be\label{gauge} %
 (2\pi\alpha')F_{t\rho}=\frac{g}{\sqrt{g^2+(2\pi\alpha')^2\rho^{2k}H^{\frac{p-q+2(k-1)}{2}}}}~,
\ee %
where the following identity has been used
\be %
 \partial_+ \left(\rho^k H^{\frac{p-q+2(k-1)}{4}} \frac{(2 \pi {\alpha'})^2
F_{t\rho}}{\sqrt{1-(2 \pi {\alpha'})^2 F_{t\rho}^2}}\right) = 0.
\ee %

It's been already explained that the sudden change in the baryon number density can locally model the collision of the heavy nuclei at RHIC. The baryon number density is defined as
$n_B=\frac{n}{N}$ where $n$ is the quark density number. In our set-up the quarks live on the $(d+1)$-dimensional defect field theory. Therefore the quark density number is $n=\frac{\tilde{n}}{Vol_{d}}$ where $\tilde{n}$ is the quark number (number of open string end
points). The time-dependent quark number can be calculated by taking the  integral of the time component of the
external source over volume space of the D$q$-brane
\be %
 \tilde{n}=\int d^{q}\xi\sqrt{-g}J^t~.
\ee %
Therefore replacing \eqref{current} in the above equation leads to
\be\label{baryon number}%
 g(x^-) = \frac{\gamma}{{\rm{Vol}}(S^k)} (2\pi)^{\frac{q-p}{2}}
 (2\pi\alpha')^{\frac{q-p+4}{2}}~ \lambda\  n_B(x^-),
\ee%
where $\lambda=g_{YM}^2 N$ and $\gamma$ is a parameter that should be fixed according to the convention. For instance in the
case of D3-D7 system \cite{Hashimoto}, $\gamma$ is $\frac{1}{4\pi}$.

So far all the calculations have been done for general values of $p$ and $q$. In order to have a stable background in the limit of zero temperature,
we consider supersymmetric systems where $q=p+4,~p+2,~p$
and $k=3,2,1$, respectively.
For such configurations we have %
\be\label{zero} %
 p-q+2(k-1)=0~.
\ee%
Therefore one can see that the $p$ and $q$ dependence in $F_{t\rho}$ and $g(x^-)$ disappears and they reduce to
\be%
 (2\pi\alpha')F_{t\rho}=\frac{g}{\sqrt{g^2+(2\pi\alpha')^2\rho^{2k}}}~,
\ee %
and %
\bea\label{maximum} %
 g(x^-)=\left\{%
\begin{array}{ll} %
    \frac{\gamma}{{{\rm{Vol}}(S^3)}} (2\pi)^{2}(2\pi\alpha')^{4}~ \lambda\ n_B(x^-), & q=p+4, \\
    \frac{\gamma}{{{\rm{Vol}}(S^2)}} (2\pi)\ (2\pi\alpha')^{3}~ \lambda \ n_B(x^-), & q=p+2, \\
    \frac{\gamma}{{{\rm{Vol}}(S^1)}} \ (2\pi\alpha')^{2}~ \lambda \ n_B(x^-), & q=p. \\
\end{array}%
\right.
\eea %

Time-dependent gauge field on the D$q$-brane modifies the metric observed by the open strings. This can be explicitly seen by using the DBI action.
We will show in the following section that this field configuration can
create an apparent horizon on the D$q$-brane which signals the
thermalization in the dual strongly coupled field theory.

\subsection{Apparent Horizon Formation}
Thermalization usually refers to an increase in the temperature of a thermodynamic system from zero to a nonzero value. We know that in the context of gauge/gravity duality the vacuum state in field theory corresponds to pure AdS space in the bulk and the thermal mixed state is dual to the AdS-black hole. So the horizon formation in the bulk can be considered as a signal of
thermalization in the dual gauge theory. In order to study the process of thermalization in a strongly coupled field theory the gauge/gravity techniques can be applied \cite{Chesler:2008hg,Bhattacharyya:2009uu}. In this framework the horizon formation usually refers to the apparent horizon formation since in contrast to the event horizon it can be obtained locally and the global knowledge of the bulk solution is not needed \cite{Chesler:2008hg}.

In the D$p$-D$q$ set-up the meson dynamics can be used to study thermalization in the field theory. The mesons are the open strings
(scalar or gauge fields) living on the D$q$-brane \cite{Kruczenski}. The D$q$-brane can
fluctuate in the directions which are \textit{transverse} or
\textit{parallel} to the background D$p$-branes. In the following
subsections we will study the dynamics of these two kinds of scalar fluctuations in the presence of a time-dependent gauge field on the D$q$-brane. It leads to calculating the modified metric on the D$q$-brane which is called the open string metric \cite{Seiberg}. Thus we can obtain the equation of the apparent horizon and compute the thermalization time-scale.

\subsubsection{Transverse Fluctuation}
Consider D$q$-brane embedding \eqref{induced metric}. The transverse fluctuations are along $\sigma$ and $\Omega_{7-p-k}$ directions. Due to the isometry $SO(8-p-k)$ we assume only $\delta\sigma$ to be nonzero. This plays the role of the scalar meson whose dynamics we would like to study. We expand the
DBI action \eqref{action} to quadratic order in
$\delta\sigma$ in the background solution $F_{t\rho}$ \eqref{gauge}. Note that since the RR field does not couple to the end point of the string on the brane it
 will not enter in the calculations for the open string metric. Therefore from now on in our calculations we neglect the CS action.
 Hence the action for the scalar meson is
\be %
 S=-\h\int dt d\rho d^dx^i d^k\theta^\alpha\sqrt{-\G}\ \G^{ab}\partial_a\delta \sigma\partial_b\delta \sigma,
\ee %
where
\be\begin{split} %
 -\G_{tt}&=\tau_q^{\frac{2}{q-1}}[1-(2\pi\alpha')^2F_{t\rho}^2]^{\frac{2-q}{1-q}}H^{\frac{p-q}{2(q-1)}},\cr
 \G_{\rho\rho}&=\tau_q^{\frac{2}{q-1}}[1-(2\pi\alpha')^2F_{t\rho}^2]^{\frac{2-q}{1-q}}H^{\frac{p+q-2}{2(q-1)}},\cr
 \G_{ij}&=\tau_q^{\frac{2}{q-1}}[1-(2\pi\alpha')^2F_{t\rho}^2]^{\frac{1}{1-q}}H^{\frac{p-q}{2(q-1)}}\delta_{ij},\cr
 \G_{\alpha\beta}&=\tau_q^{\frac{2}{q-1}}[1-(2\pi\alpha')^2F_{t\rho}^2]^{\frac{1}{1-q}}\rho^2H^{\frac{p+q-2}{2(q-1)}}\hat{g}_{\alpha\beta}.
\end{split}\ee %
$i=1,..,d$ is the spatial directions of the $(d+1)$-dimensional
defect. $\theta^\alpha$ ($\alpha=1,..,k$) is the angular variable
on $S^k$ and $\hat{g}_{\alpha\beta}$ is the metric on the unit
$k$-sphere. The surface of the presumably apparent horizon, which is
defined locally as a surface whose area variation vanishes along the
null ray which is normal to the surface, is given by
\footnote{One can see that for the supersymmetric
cases the apparent
horizon surface area simplifies to
$$ V_{\rm{surface}}=V_d {\rm{Vol}}(S^k) \frac{\tau_q}{2\pi\alpha'} H^{\frac{1}{2}} \sqrt{g^2 + (2 \pi \alpha')^2 \rho^{2k}}.$$ }
\be\begin{split}\label{surface} %
 V_{\rm{surface}}&=\int
 d^dx^id^k\theta^\alpha\Big(\prod_{i=1}^d\G_{ii}\prod_{\alpha=1}^{k}\G_{\alpha\alpha}\Big)^\h\cr
 &=V_d {\rm{Vol}}(S^k) \frac{\tau_q}{2\pi\alpha'} H^{\frac{1}{2}} \sqrt{g^2 + (2 \pi \alpha')^2 \rho^{2k} H^{\frac{p-q+2(k-1)}{2}}},
\end{split}\ee %
where this should satisfy %
\be\label{apparent}%
 dV|_{dt=-dz} = 0.
\ee%
Therefore the equation for the apparent horizon reduces to
\be\begin{split} %
 \label{master}
 4gg'H^\h
 \rho+(p-7)g^2+\nu(2\pi\alpha')^2H^{\frac{p-q+2(k-1)}{2}}\rho^{2k}=0,
\end{split}\ee %
where $\nu=\frac{1}{2}\left[(p-7)(p-q+2k)+4k\right]$. We call this
the \textit{master equation}. For supersymmetric cases, the master
equation can be simplified further
\be\begin{split} %
 \label{master}
 4gg'H^\h
 \rho+(p-7)g^2+ (p-7+2k) (2\pi\alpha')^2\rho^{2k}=0.
\end{split}\ee %
Since working with $z$ coordinate is more appropriate we replace $\rho$ with $z$ in \eqref{master} and get
\be\label{master1}%
2 (5-p) g g' z + (p-7) g^2 + (2\pi\alpha')^2 (p-7+2k)
(\frac{5-p}{2})^{\frac{4k}{p-5}} z^{\frac{4k}{p-5}}
R^{\frac{2k(p-7)}{p-5}}=0,
\ee%
where the length scale $R$ can be written in terms of the t'Hooft
coupling constant $\lambda$ as %
\be\label{length scale} %
 R^{\frac{p-7}{p-5}}=\sqrt{2}(2\pi\alpha')(2\pi)^{\frac{p+1}{2(p-5)}}\ \Gamma(\frac{7-p}{2})^{\frac{1}{5-p}}\ \lambda^{\frac{1}{5-p}}. %
\ee%
The formula \eqref{master1} is the main equation that will be used to calculate the thermalization time for the transverse fluctuations.

\subsubsection{Parallel Fluctuation}
As it was mentioned earlier in addition to the transverse fluctuations of the D$q$-brane one can also study the ones
 which are parallel to the background
D$p$-branes. These parallel fluctuations exist only for $q=p+2$ and $p$. This can be easily seen from the configuration (\ref{graph}). Note that for $q=p+2$ we have $d=p-1$ and the only parallel fluctuation is along $p$ direction. For $q=p$ there are two directions for parallel fluctuations, $p-1$ and $p$, and the fundamental matter lives on the spatial $(d=p-2)$-dimensional defect.

Similar to the transverse fluctuations the DBI action reduces to
\be %
 S=-\h \int dt d\rho d^dx^i d^k\theta^\alpha \sqrt{-\G}\ \G^{ab}\partial_a\delta x^n
 \partial_b\delta x^n~,
\ee %
where $\delta x^n$ represents small fluctuation parallel to
D$p$-branes. The components of the metric are %
\be\begin{split} %
\label{metricparallel}
 -\G_{tt}&= \tau_q^{\frac{2}{q-1}}[1-(2\pi\alpha')^2F_{t\rho}^2]^{\frac{2-q}{1-q}}H^{\frac{p-q-4}{2(q-1)}},\cr
 \G_{\rho\rho}&= \tau_q^{\frac{2}{q-1}}[1-(2\pi\alpha')^2F_{t\rho}^2]^{\frac{2-q}{1-q}}H^{\frac{p+q-6}{2(q-1)}},\cr
 \G_{ij}&=\tau_q^{\frac{2}{q-1}}[1-(2\pi\alpha')^2F_{t\rho}^2]^{\frac{1}{1-q}}H^{\frac{p-q-4}{2(q-1)}}\delta_{ij},\cr
 \G_{\alpha\beta}&=\tau_q^{\frac{2}{q-1}}[1-(2\pi\alpha')^2F_{t\rho}^2]^{\frac{1}{1-q}}\rho^2H^{\frac{p+q-6}{2(q-1)}}\hat{g}_{\alpha\beta}.
\end{split}\ee %
Therefore we can now calculate the equation for the apparent horizon area\footnote{In
supersymmetric cases we have
 $$V_{\rm{surface}}=V_d {\rm{Vol}}(S^k) \frac{\tau_q}{2\pi\alpha'}  H^{-\frac{1}{2}} \sqrt{g^2 + (2 \pi \alpha')^2 \rho^{2k}}.$$ } %
\be%
 V_{\rm{surface}}=V_d {\rm{Vol}}(S^k) \frac{\tau_q}{2\pi\alpha'} H^{-\frac{1}{2}} \sqrt{g^2 + (2 \pi \alpha')^2 \rho^{2k} H^{\frac{p-q+2(k-1)}{2}}}.
\ee%
Similar to the previous subsection we apply the condition \eqref{apparent} and the master equation for the parallel fluctuations reduces to
\be%
\label{masterp}
 2 (5-p) g g' z - (p-7) g^2 - (2\pi\alpha')^2 (p-7-2k)
 (\frac{5-p}{2})^{\frac{4k}{p-5}} z^{\frac{4k}{p-5}}
 R^{\frac{2k(p-7)}{p-5}}=0,
\ee%
where the length scale $R$ was introduced in \eqref{length scale}. The master equation \eqref{masterp} is very similar to the one obtained for transverse fluctuations \eqref{master1} except for a couple of sign differences. Note that we have calculated the master equation for the supersymmetric cases.

Before closing this section we would like to emphasize on the
observation that for the general values of $q$, not necessarily
supersymmetric ones, the only change in the equations
\eqref{master1} and \eqref{masterp} appears in the power of $z$, in
their last term. Since $p-q+2(k-1)$ is not zero any more in the
non-supersymmetric brane configurations the power of $z$,
$\frac{4k}{p-5}$, is replaced by $\frac{p-q+2(3k-1)}{p-5}$. Note
that this is a rough argument and we have ignored discussing the
stability of the backgrounds. We will see from our results in the following section
that this indicates that the thermalization time scale
behaviour does not rely on supersymmetry.

\section{Baryon Injection Model}
To be able to determine the time-scale of thermalization, the source
function $g(x^-)$ should be specified. Similar to the function
considered in \cite{Hashimoto}, we assume
\bea %
\label{g}
 g(x^-)=\left\{%
\begin{array}{ll}
    0 & x^-<0 \\
    g_{max}\ \omega x^- & 0<x^-<\frac{1}{\omega} \\
    g_{max} & \frac{1}{\omega}<x^- \\
\end{array}%
\right. \eea %
Using the equation \eqref{baryon number} it shows that
the baryon density $n_B$ is zero in the beginning which resembles the situation before the collision of heavy ions. At $x^- = 0$  it starts changing to reach a maximum value at $x^-=\frac{1}{\omega}$ and remains constant since then. The maximum value of
source function $g_{max}$ is given by \eqref{maximum} where $n_B(x^-)$ is replaced by its constant maximum value, $n_B$
\be\label{gmax}%
 g_{max} = \frac{\gamma}{{\rm{Vol}}(S^k)} (2\pi)^{\frac{q-p}{2}}
 (2\pi\alpha')^{\frac{q-p+4}{2}}~ \lambda\  n_B.
\ee%
This is similar to the situation where the baryonically-charged heavy ions approach each other and the baryon number chemical potential changes locally.

The time-dependence of the baryon number can be chosen to have a
general form, $(\omega x^-)^n$, where $n\geq 0$\footnote{If we assume $g(x^-)$ to
scale as $g_{max} (\omega x^-)^n$ the change in the thermalization time appears in the overall power $\beta$
and the power of $\omega$ as $$t_{th}\sim(\frac{\lambda^\alpha}{n_B^2 \omega^{2n}})^\beta.$$ Note that, in contrast to transverse fluctuations, $n=0$ does not give a real solution for parallel ones. We will see this in subsection \ref{subsec:PF}.}. This is a dimensionless combination and
guarantees that $g(x^-)$ gets its maximal
value at $x^-=\frac{1}{\omega}$.

We consider the supersymmetric configurations which fall into these
categories:

\subsection{(p,p + 4) system} 
\label{subsec:p+4}
In this set-up of brane configuration we consider $q$ to be $p+4$ where $0 \leq p <5$. As it's been shown in (\ref{graph}) there is no parallel fluctuations in this case and one needs to consider only the transverse fluctuations.

\subsubsection{Transverse Fluctuation}
If we define $g(x^-)=g_{max}~y(x^-)$ where $g_{max}$ is given by
\eqref{gmax} and set $k$ to be 3, the master equation \eqref{master1} becomes %
\be\begin{split} %
 (p&-7)y^2+2(5-p)y y' z \cr
 &+\frac{1}{8\gamma^2}(p-1)(\frac{5-p}{2})^{\frac{12}{p-5}}(2\pi)^{\frac{3(p+1)}{p-5}}\ \Gamma(\frac{7-p}{2})^{\frac{6}{5-p}}
 \frac{\lambda^{\frac{2(p-2)}{5-p}}}{n_B^2}z^{\frac{12}{p-5}}=0.
\end{split}\ee %
To warm up, we start with the D4-D8 brane configuration.

\begin{figure}[ht]
\begin{center}
\includegraphics[width=2.1 in]{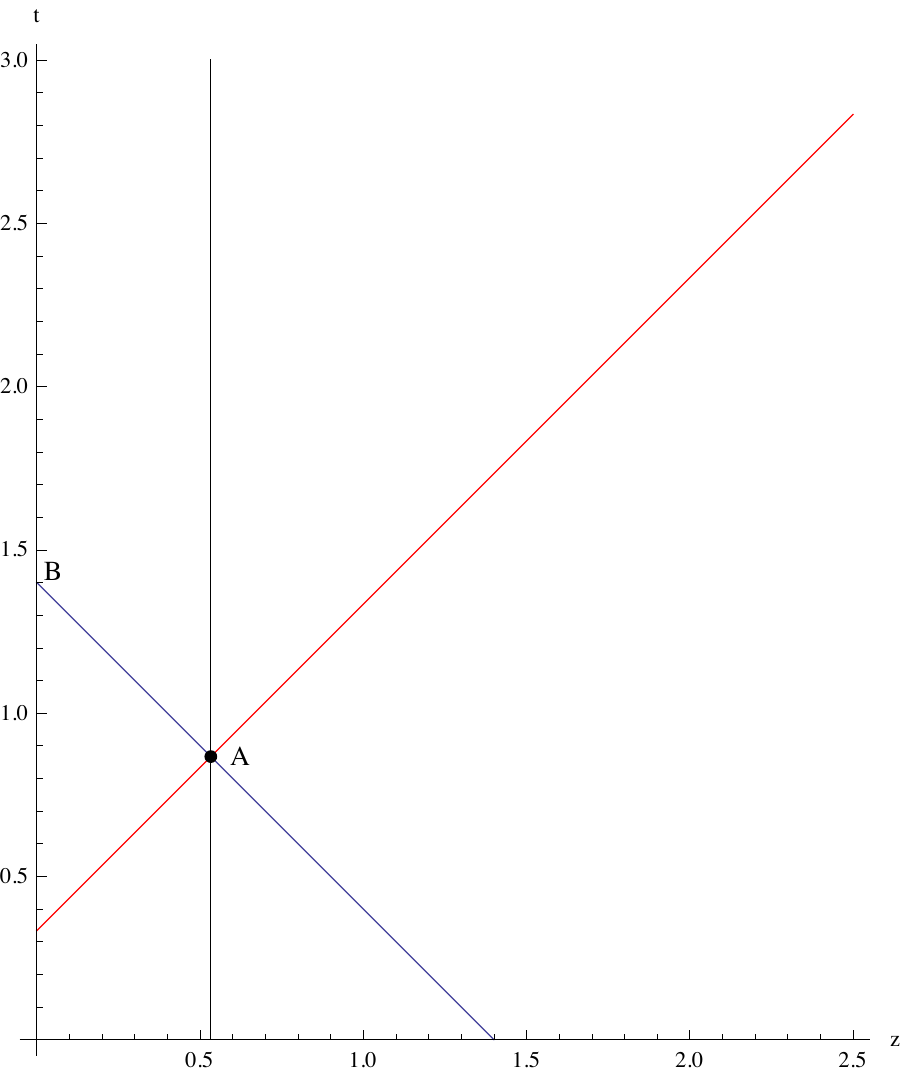}
\caption{The location of the apparent horizon for $t>z+\frac{1}{\omega}$ in the $t-z$ plane. The black, red and blue lines represent \eqref{timesacle2}, $t=z+\frac{1}{\omega}$ and the earliest null ray which reaches the boundary. The parameters are chosen as $\omega=3$
and $(\frac{\lambda^2}{\gamma n_B})^\frac{1}{6}=\frac{10}{\omega}$
for D4-D8 system.\label{d4d81}}
\end{center}
\end{figure}

\begin{description}
    \item[D4-D8]
\end{description}
 The master equation is then %
\be %
\label{master48}
 \frac{3\lambda^4}{4096 \pi ^{12}\gamma^2 n_B^2 z^{12}}-3 y^2+2 z y
 y'=0.
\ee %
The location of the apparent horizon on the D8-brane is obtained by solving the above equation and determine $t$ in terms of $z$. We get
\bse %
\begin{align}
    \label{timesacle1}t&=\frac{4z}{3}+\frac{1}{192}\sqrt{4096z^2+\frac{9\lambda^4}{z^{12}\pi^{12}\gamma^2n_B^2\omega^2}}~,\ \ \ t<z+\frac{1}{\omega},\\
    \label{timesacle2}z&=0.16(\frac{\lambda^2}{\gamma n_B})^{\frac{1}{6}},~\hspace{4.70cm} t>z+\frac{1}{\omega}.
\end{align} %
\ese %
\begin{figure}[ht]
\begin{center}
\includegraphics[width=2.8 in]{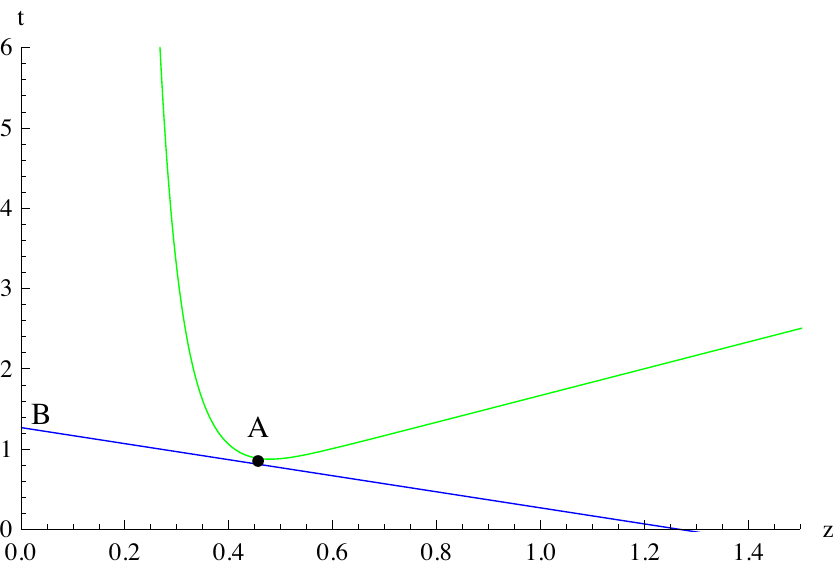}
\caption{The location of the apparent horizon for $t<z+\frac{1}{\omega}$ in the $t-z$ plane. The green curve represents \eqref{timesacle1} and the blue line shows the tangent null ray. The parameters are chosen as $\omega=0.5$ and
$\frac{\lambda^4}{\gamma^2n_B^2}=\frac{1}{\omega^{12}}$ for D4-D8
system.\label{d4d8}}
\end{center}
\end{figure}%
The curve \eqref{timesacle1} and the line \eqref{timesacle2} are plotted in figures \ref{d4d8} and \ref{d4d81} in green and black colours, respectively. The goal of these calculations is to compute the thermalization time in the boundary field theory. This in fact is the time that a boundary observer starts to see the apparent horizon formation on the D8-brane. For $t<z+\frac{1}{\omega}$ the null ray which is tangent to \eqref{timesacle1} is the earliest one that conveys the information to the
boundary. It has been shown by the blue line in figure \ref{d4d8}. The other null rays which cross \eqref{timesacle1} reach the boundary at later
times. The tangent point A coordinates are
\be %
 z_A=0.46(\frac{\lambda^2}{\gamma n_B\omega})^\frac{1}{7},  \ \ \ t_A=0.81(\frac{\lambda^2}{\gamma n_B\omega})^\frac{1}{7}.
\ee %
So the thermalization time is
\be %
 t_{th}=t_A+z_A\sim(\frac{\lambda^2}{n_B\omega})^\frac{1}{7}.
\ee %
An interesting observation is that the thermalization time depends only on $\lambda$ and the baryon injection parameters $n_B$ and $\omega$.

Similar argument can be made for \eqref{timesacle2}. In that case the thermalization time is determined by the point A coordinates
in figure \ref{d4d81}. Point A is where the line \eqref{timesacle2} crosses the ingoing null ray $t=z+\frac{1}{\omega}$, red line in that figure.
So the null ray which passes through this point and reaches the boundary gives the earliest time that a boundary observer
sees the horizon formation. This happens especially if %
\be\label{inequality} %
 (\frac{\lambda^2}{n_B\omega})^\frac{1}{7}\gg\frac{1}{\omega}~,
\ee %
and thermalization time becomes %
\be %
 t_{th}= 2z_A+\frac{1}{\omega}~,
\ee %
where %
\bse\begin{align}
 t_A&=z_A+\frac{1}{\omega}~, \\
 z_A&=0.16(\frac{\lambda^2}{\gamma n_B})^{\frac{1}{6}}.
\end{align}\ese %
Inequality \eqref{inequality} then indicates %
\be %
 t_{th}\sim(\frac{\lambda^2}{n_B})^{\frac{1}{6}}.
\ee %

\subsubsection*{Dp-D(p+4)}
The above calculation can be generalized to other values of $p$. We conclude that
\begin{itemize}
\item  $0\leq p(\neq 1)<5$
           \be %
            t_{th} \sim \left(\frac{\lambda^{\frac{2(p-2)}{5-p}}}{n_B^2\omega^2}\right)^{\frac{5-p}{2(11-p)}},
           \ee %
and if $\left(\frac{\lambda^{\frac{2(p-2)}{5-p}}}{n_B^2\omega^2}\right)^{\frac{5-p}{2(11-p)}}\gg\frac{1}{\omega}$  we have %
           \be %
            t_{th} \sim \left(\frac{\lambda^{\frac{2(p-2)}{5-p}}}{n_B^2}\right)^{\frac{5-p}{12}}.
           \ee %

\item  $p=1$

Setting $p=1$ the system behaves differently from the other values of $p$. This is due to the fact that at $p=1$ one of the terms in the master equation vanishes and this trivializes the calculations.
\end{itemize}


\subsection{(p,p + 2) system} 
Consider a D$(p+2)$-brane supersymmetrically embedded in the background obtained from the near horizon geometry of $N$ coincident D$p$-branes where $1 \leq p <5$. $k$ equals 2 and the dual gauge theory is $(p+1)$-dimensional, but the fundamental hypermultiplet has been introduced on a  $(d+1=p)$-dimensional defect. There are three and one set of fluctuations along transverse and parallel directions, respectively. In the following subsections we will discuss them individually.

\subsubsection{Transverse Fluctuations}
For this configuration the master equation \eqref{master1} reduces to
\be\label{masterp+2}\begin{split} %
 2(5&-p)y y'z+(p-7)y^2\cr
 &+\frac{16}{9}(p-3)(2\pi)^{\frac{2(p+1)}{p-5}}(\frac{5-p}{2})^{\frac{8}{p-5}}
 \ \Gamma(\frac{7-p}{2})^{\frac{4}{5-p}}\frac{\lambda^{\frac{2(p-3)}{5-p}}z^{\frac{8}{p-5}}}{\gamma^2 n_B^2}=0.
\end{split}\ee %
If we repeat the same calculations done in
the previous section, the thermalization time for arbitrary values of
$p$ is
\begin{itemize}
\item $1\leq p(\neq3)<5$
          \be %
          \label{resultp+2t}
         t_{th} \sim \left(\frac{\lambda^{\frac{2(p-3)}{5-p}}}{n_B^2\omega^2}\right)^{\frac{5-p}{2(9-p)}},
          \ee %
and if the time scale lies in the limit where $\left(\frac{\lambda^{\frac{2(p-3)}{5-p}}}{n_B^2\omega^2}\right)^{\frac{5-p}{2(9-p)}}\gg\frac{1}{\omega}$, we obtain %
           \be %
            t_{th} \sim \left(\frac{\lambda^{\frac{2(p-3)}{5-p}}}{n_B^2}\right)^{\frac{p-5}{8}}.
           \ee %
\item $p=3$

Note that similar to $p=1$ case in the D$p$-D$(p+4)$ system, choosing $p=3$ sets one of the terms in the master equation to zero and the thermalization time can not be obtained. Interestingly we will see in the following that the parallel fluctuation produces the result we expect for $p=3$.
\end{itemize}

\subsubsection{Parallel Fluctuations}
\label{subsec:PF}
As it was mentioned earlier there is only one parallel fluctuation for D$p$-D$(p+2)$ system which is along the $x^p$ direction. The master equation \eqref{masterp} for $k=2$ reduces to
\be%
\label{masterp+2p}
 2 (5-p) g g' z - (p-7) g^2 - (2\pi\alpha')^2 (p-11)
 (\frac{5-p}{2})^{\frac{8}{p-5}} z^{\frac{8}{p-5}}
 R^{\frac{4(p-7)}{p-5}}=0~.
\ee%
Let us start with an explicit example.
\begin{description}
    \item[D4-D6]
\end{description}
Setting $p=4$ the apparent horizon equation becomes
\be %
\label{master46}
 3 y^2+2 z y y'+\frac{7\lambda^2}{36\pi ^{8}\gamma^2 n_B^2 z^{8}}=0.
\ee%
This looks similar to the master equation for D4-D8 brane system \eqref{master48}. We follow the same analogy here.  The solution to the above equation which specifies the location of the apparent horizon reads
\be %
\label{first} t=\frac{2z}{3}+\frac{1}{3}\sqrt{z^2-\frac{7\lambda^2}{12z^{8}\pi^{8}\gamma^2n_B^2\omega^2}}~,\
 \ \ \  t<z+\frac{1}{\omega}.
\ee %
Note that there is no real solution for $z$ when $t>z+\frac{1}{\omega}$. Thus the thermalization time is only fixed by the solution \eqref{first}.
For $t<z+\frac{1}{\omega}$ the expression under the square root must be positive. Fortunately we can always find a solution for $z$ that respects this condition. The coordinates of the tangent point A to the curve \eqref{first} will determine the thermalization time. This has been fully discussed in D4-D8 case so we state the result. The coordinates of the tangent point A are
\be\begin{split}
 z_A=\frac{0.39}{\gamma^{1/5}}(\frac{\lambda}{n_B\omega})^{1/5},\cr
 t_A=\frac{0.45}{\gamma^{1/5}}(\frac{\lambda}{n_B\omega})^{1/5}.
\end{split}\ee %
Hence the thermalization time on the boundary reads
\be %
 t_{th}=t_A+z_A\sim (\frac{\lambda}{n_B\omega})^{1/5}.
\ee %
If we set $p=4$ in the result for the transverse fluctuation \eqref{resultp+2t} we get the same thermalization time scale, up to a numeric constant. It is interesting that although the apparent horizon equations are different for transverse and parallel fluctuations their thermalization times observed on the boundary exactly resemble each other.

\begin{description}
    \item[Dp-D(p+2)]
\end{description}
The above result can be generalized to other values of $p$ and we get
\be %
\label{resultp+2p}
 t_{th} \sim \left(\frac{\lambda^{\frac{2(p-3)}{5-p}}}{n_B^2\omega^2}\right)^{\frac{5-p}{2(9-p)}}~.
\ee %
 One can conclude that  the thermalization time for parallel fluctuations and transverse ones \eqref{resultp+2t} look completely alike for various values of $p$. An interesting observation is that the time scale \eqref{resultp+2p} works fine for $p=3$ case in contrast to the transverse one. We can see that for $p=3$ there is no dependence on the t'Hooft coupling $\lambda$ in the thermalization time scale.

\subsection{(p,p) system}
 Consider the brane configuration where $q=p$. In order to supersymmetrically embed one D$p$-brane in the background geometry of $N$ D$p$-branes we choose $2 \leq p < 5$. As it was mentioned earlier the fundamental hypermultiplet is confined to a $(d+1=p-1)$-dimensional surface.  $k=1$ for this set-up. Therefore there are two sets of both, parallel and transverse fluctuations which we will elaborate more on them in the following.

\subsubsection{Transverse Fluctuations}
The master equation \eqref{master1} in this case reduces to
\be\begin{split} %
 (p&-7)y^2+2(5-p)y y' z \cr
 &+\frac{1}{2\gamma^2}(p-5)(\frac{5-p}{2})^{\frac{4}{p-5}}(2\pi)^{\frac{3(p-3)}{p-5}}\ \Gamma(\frac{7-p}{2})^{\frac{2}{5-p}}
 \frac{\lambda^{\frac{2(p-4)}{5-p}}}{n_B^2}z^{\frac{4}{p-5}}=0.
\end{split}\ee %
The thermalization time-scale for general values of $p$ becomes%
\begin{itemize}
\item  $2\leq p<5$
           \be %
            t_{th} \sim \left(\frac{\lambda^{\frac{2(p-4)}{5-p}}}{n_B^2\omega^2}\right)^{\frac{5-p}{2(7-p)}},
           \ee %
and in the limit where $\left(\frac{\lambda^{\frac{2(p-4)}{5-p}}}{n_B^2\omega^2}\right)^{\frac{5-p}{2(7-p)}}\gg\frac{1}{\omega}$, we have %
           \be %
            t_{th} \sim \left(\frac{\lambda^{\frac{2(p-4)}{5-p}}}{n_B^2}\right)^{\frac{p-5}{4}}.
           \ee %
\end{itemize}

\subsubsection{Parallel Fluctuation}
For D$p$-D$p$ system there are two directions for parallel fluctuations. The master equation \eqref{masterp} reads
\be\begin{split} %
 -(p&-7)y^2+2(5-p)y y' z \cr
 &-\frac{1}{2\gamma^2}(p-9)(\frac{5-p}{2})^{\frac{4}{p-5}}(2\pi)^{\frac{3(p-3)}{p-5}}\ \Gamma(\frac{7-p}{2})^{\frac{2}{5-p}}
 \frac{\lambda^{\frac{2(p-4)}{5-p}}}{n_B^2}z^{\frac{4}{p-5}}=0.
\end{split}\ee %
Interestingly we observe that similar to D$p$-D$(p+2)$ system, the thermalization time for parallel fluctuations has the same scaling behaviour as the transverse fluctuations. Therefore for general $p$ we obtain
\begin{itemize}
\item  $2\leq p<5$
           \be %
            t_{th} \sim \left(\frac{\lambda^{\frac{2(p-4)}{5-p}}}{n_B^2\omega^2}\right)^{\frac{5-p}{2(7-p)}}.
           \ee %
\end{itemize}

\subsection{Concluding Remarks}
We conclude this section mentioning some interesting observations. It's been already mentioned that the thermalization time for parallel and transverse fluctuations are the same, up to a numeric coefficient, even though the apparent horizon equations are different. The theories we discussed here are not necessarily conformal and their dual background geometries are not AdS. But the time-scale behaviour still gets the same form as an AdS background with a conformal field theory dual such as D3-D7 system. Even for general values of $q$ for which the solution is not supersymmetric the general form of the thermalization time \eqref{universal} persists. This seems to approve the claim that the thermalization time-scale behaves universally. But one may ask to what extent this discussion works. We leave it as an open question.

Let us consider the case where $p=3$. As it is well known the dual gauge theory is ${\cal N}=4$ SYM theory which lives on the 4-dimensional boundary of AdS space. If we require to preserve supersymmetry we can add three types of probe flavour branes to this background. They are D3, D5, and D7-branes. Adding these branes will modify the dual gauge theory to ${\cal N}=4$ coupled to ${\cal N}=2$ fundamental
hypermultiplet. The dynamical quarks live on $(1+1),~(2+1)$ and $(3+1)$-dimensional defects for $q=3,~5$ and $7$, respectively. Interestingly we see
that although the 4-dimensional gauge theory is the same for all of these brane set-ups these theories produce different
time scales for thermalization which are
\bea %
 t_{th}\sim\left\{%
\begin{array}{ll}
    (\frac{\lambda}{n_B^2 \omega^2})^{\frac{1}{8}}, & {\rm{D3-D7}}, \\
    \\
    (\frac{1}{n_B^2 \omega^2})^{\frac{1}{6}}, & {\rm{D3-D5}}, \\
    \\
    (\frac{1}{n_B^2 \omega^2 \lambda})^{\frac{1}{4}}, & {\rm{D3-D3}}. \\
\end{array}%
\right. \eea %
Therefore we can conclude that the thermalization time scales differently with the t'Hooft coupling, $\lambda$. The main difference between these theories come from the fact that the dynamical quarks are confined to defects with different dimensions.

Another intriguing observation is when we consider the dimension of the defect to be $(2+1)$. The brane configurations to get it are D2-D6, D3-D5 and D4-D4. One can see that in all of these cases the dependence on $\lambda$ in the thermalization time disappears. It is very interesting that for this specific choice all the brane configurations give the same value for the thermalization time which is $(\frac{1}{n_B^2 \omega^2})^{\frac{1}{6}}$. Note that this doesn't happen in the other dimensions for the defect such as $(1+1)$ and $(0+1)$ dimensions.

In this paper we have studied thermalization for the scalar mesons. One can generalize this calculation and study the dynamics of the vector mesons. Moreover the universal behaviour of the thermalization time-scale can be investigated in other gravity backgrounds such as anisotropic Lifshitz ones. It is also interesting to study the time-scale of the change in the temperature if one starts from thermal field theory, dual to black hole background, instead of zero-temperature one and then injects energy into it. We postpone these to the future works.

\section*{Acknowledgement }
We would like to thank N. Abbasi, M. Alishahiha, A. Davody and M. M. Sheikh-Jabbari for fruitful comments and discussions.

\end{document}